# Sensitivity of coherent anti-Stokes Raman lineshape to time asymmetry of laser pulses

Michele Marrocco,[1,*] Emil Nordström,[2] Per-Erik Bengtsson,[2]
[1]ENEA, via Anguillarese 301, 00167 Rome, Italy
[2]Combustion Physics, Lund University, SE-22100, Sweden
*Corresponding author: michele.marrocco@enea.it

We show that coherent anti-Stokes Raman lineshapes do not follow known spectral profiles if the time asymmetry of realistic laser pulses is taken into account. Examples are given for nanosecond and picosecond laser pulses commonly employed in frequency-resolved coherent anti-Stokes Raman scattering. More remarkably, the analysis suggests an effect of line narrowing in comparison to the customary approach, based primarily on the Voigt lineshape.

## Introduction

Coherent anti-Stokes Raman scattering (CARS) is one of the fundamental spectroscopic tools of nonlinear optics [1-4]. It involves the use of three laser fields and, among several applications, gas analysis with nanosecond (ns) and picosecond (ps) lasers has attracted a lot of interest over the years [3, 5-8]. Within this context, the diagnostic capacity depends critically on the physical understanding of CARS spectra and, for this reason, one has to secure the information contained in the third-order nonlinear susceptibility describing the emergence of an isolated CARS line. This optical response function is very well known and is usually represented in the frequency domain as a fourth-rank tensor of the kind

$$\chi^{(3)}_{ijkl}(\Delta) = \frac{K_{ijkl}}{\Delta - i\Gamma/2} \tag{1}$$

where $K_{ijkl}$ contains the information about the Raman molecule (i.e., Raman frequency, Boltzmann distribution of the molecular population, polarizability derivatives and other molecular parameters), $\Delta$ is the Raman detuning and $\Gamma$ is the Raman FWHM linewidth whose value is of paramount importance. It is then tempting to conclude that Eq. (1) leads to Lorentzian profiles in CARS spectra where single Raman modes are isolated.

The conclusion is, however, subject to constraints. In particular, disregarding the interference caused by the non-resonant background, the deviation of the CARS lineshape from the true Lorentzian profile of Eq. (1) has been discussed by several authors with reference to the use of multi-mode lasers [9-13]. On the other hand, the advent of the injection-seeding (or pulsed injection locking) technology [14-17] has made possible to realize single-mode Q-switched laser operation resulting in near-transform-limited pulses with spectral bandwidths $\delta\tilde{\nu}$ that are smaller than the typical Raman widths $\Gamma$.

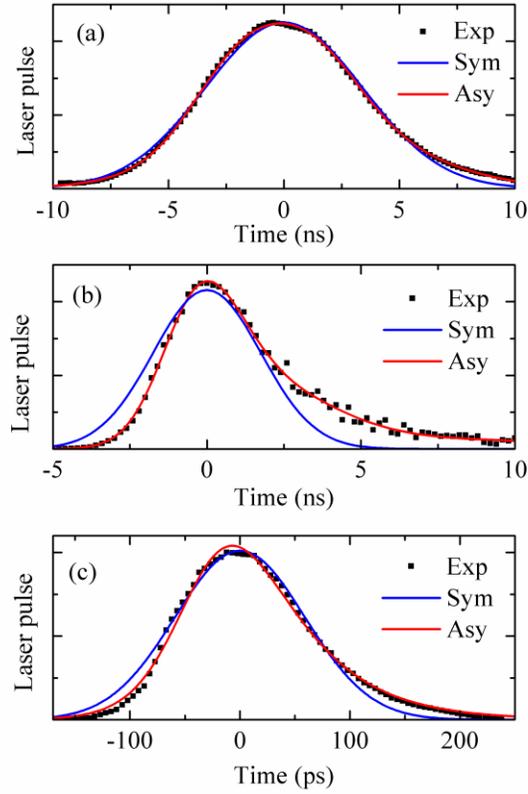

Fig. 1. Experimental laser pulses with their symmetric and asymmetric simulations. Gaussian profiles are used to simulate symmetric pulses. The square of the asymmetric hyperbolic secant is used to model the asymmetry.

For such laser systems, the reference pulse assumed in most of the CARS studies has a Gaussian spectral envelope and this results in the traditional Voigt lineshape obtained as real part of the complex error function [9, 13, 18-20]. Beyond its undoubted importance in laser spectroscopy at large, the Voigt function is at the core of various fitting codes developed to study numerically the complex fine structures of CARS spectra. The most known is the so-called CARSFT code [21], but others have been published [20, 22].

### Role of time asymmetry in an isolated CARS lineshape

In this work, we demonstrate that realistic temporal aspects of the laser pulses could determine a significant change in the CARS lineshape in comparison to the Lorentz and, most importantly, Voigt profiles. For instance, Gaussian functions to approximate spectral envelopes of the electric fields of transform-limited laser pulses contrast against the time asymmetry of realistic laser pulses [15-17, 23-27] and this inconsistency has some consequences for the traditional CARS theory.

The proof is based on the comparison of CARS lineshapes produced by means of the laser pulses that are shown in Fig. 1. The solid symbols represent real transform-limited Nd:YAG laser pulses that are common to many gas-phase CARS applications and whose temporal profiles are chosen according to the following criteria. First of all, in Fig. 1(a), the ideal situation of a very small asymmetry is exemplified for a laser pulse of about 8 ns. This pulse is ideal in that the symmetric fit (blue line), based on the Gaussian envelope, differs very slightly from the asymmetric dependence (red line) used to simulate the real pulse. By contrast, Fig. 1(b) is relative to a shorter pulse with a considerable asymmetry and, for this reason, represents the opposite extreme of Fig. 1(a). However, for the sake of argument, we include the case of a moderate time asymmetry that is typical of ps-CARS [28-30]. This case is visualized in Fig. 1(c) for a laser pulse of about 140 ps. The symmetric dependence for the fits of Fig. 1(b) and (c) is assumed Gaussian (blue line). The much more accurate asymmetric fits (red line) are instead characterized by the asymmetric hyperbolic secant that is often used to model realistic laser pulses [27, 31-33].

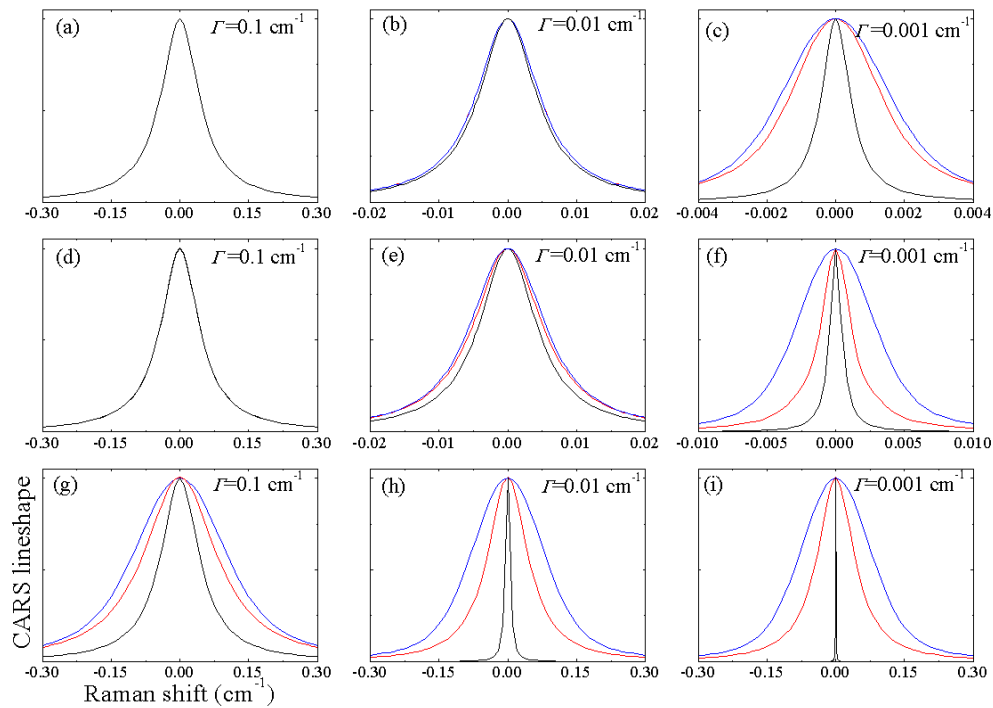

Fig. 2. CARS lineshapes related to the laser pulses of Fig. 1 [plots in the upper row are relative to Fig. 1(a), plots in the middle row are relative to Fig. 1(b) and plots in the lower row are relative to Fig. 1(c)]. The black line represents the Lorentzian profile of Eq. (1) with Raman linewidths $\Gamma$ indicated at the top of each graph. The Voigt CARS lineshape produced by the symmetric Gaussian laser pulses is represented by the blue line. The red line shows instead the CARS lineshape associated with the asymmetric laser pulse.

Given the laser pulses of Fig. 1, we now examine the corresponding CARS lineshapes obtained for the traditional laser arrangement of degenerate pump and probe pulses mixed with a broadband Stokes laser within a Raman gas medium of linewidth $\Gamma$ whose values are chosen between the extremes of 0.1 and 0.001 cm$^{-1}$. Physically, this range corresponds to moderate and very low gas pressures with variable temperatures.

The results for the laser pulses of Fig. 1 are shown in Fig. 2, where the CARS lineshapes (red and blue lines) are compared to the Lorentzian line of Eq. (1) (black line). The figure is organized as follows. The plots in the upper row refer to the case of the ideal laser pulse of Fig. 1(a). In the middle row, we find the CARS lineshapes relative to the strong time asymmetry of Fig. 1(b) and, finally, the bottom row of Fig. 2 reports the CARS lineshapes consequent on the moderate asymmetry of the ps laser pulse of Fig. 1(c).

Starting from the top of Fig. 2, we notice that the CARS lineshape is indistinguishable from the Lorentzian line [square modulus of Eq. (1)] when a large Raman linewidth is considered, that is $\Gamma = 0.1 \, \text{cm}^{-1}$ in Fig. 2(a). This was expected. Less expected is the functional dependence of the CARS lineshape that departs from the Lorentzian limit for the small Raman linewidth of 0.01 cm$^{-1}$ in Fig. 2(b). Despite this deviation, we cannot appreciate differences between the Voigt line of the symmetric pulse and the spectral line of the asymmetric fit of Fig. 1(a). The differences appear instead in the limit of very low Raman width ($\Gamma = 0.001 \, \text{cm}^{-1}$) of Fig. 2(c). In particular, the CARS lineshape of the asymmetric pulse is narrower than the Voigt lineshape of the symmetric pulse.

Similar to the previous result for $\Gamma = 0.1 \, \text{cm}^{-1}$, the strong asymmetry of Fig. 1(b) does not affect the CARS lineshape of Fig. 2(d) and, in this limit, the time asymmetry of ns laser pulses seems to be negligible. But, as soon as the Raman linewidth decreases, the time asymmetry plays a fundamental role and introduces again an effect of line narrowing as shown more convincingly in Fig. 2(f) for the extreme of a very low Raman linewidth. The line narrowing takes place simply because of the narrower spectral laser line of the transform-limited asymmetric pulse in comparison to its symmetric analog. The effect becomes even more manifest with the ps pulses of Fig. 1(c). Indeed, the corresponding results of Figs. 2(g), (h) and (i) are all consistent with a relevant role played by the time asymmetry that has never been considered in the recent literature of ns- and ps-CARS [7, 28-30, 34-36] apart from wildly asymmetric ps pulses characterized by a pure exponential decay [37].

To recapitulate, the time asymmetry of laser pulses introduces an effect of line narrowing if compared to the Voigt CARS lineshape. The onset of such an effect depends on the Raman linewidth and is of course more intense with ps laser pulses that have broader spectral bandwidth. It is then germane to the discussion to point out that the effect was found in high-resolution ns-CARS measurements of $N_2$ and $O_2$ [38]. The authors observed that the experimental spectra required that the measured Q-branch lineshapes of both species had a narrower peak at the line center than the best-fit Voigt profile. However, the CARS sensitivity to such a line narrowing was not investigated and the question has remained open.

**Application to N2 CARS spectra**

Next, we adapt the comparative analysis of Fig. 2 for the peculiarity of $N_2$ CARS that is often use in gas thermometry [3-8]. The following simulation refers to pure nitrogen described by the isolated-line approach with collisional narrowing adjusted to the atmospheric pressure and a temperature of 2400 K. In addition, we disregard the non-resonant contribution that has nothing to do with the matter of our argument. The result relative to the ps laser pulses of Fig. 1 is shown in Fig. 3. For the purposes of this comparison between lineshapes, the curves are normalized to their own maximum and, by doing so, some mismatches are recognizable. First of all, significant differences among signal levels are found near the bandhead at 2330 cm$^{-1}$. In addition, different peak values are found in the hot band and especially around 2292 cm$^{-1}$. The distinctive feature of Fig. 3 is however shown in the inset where the effect of line narrowing is demonstrated for the rotational transitions at $J = 19$, 20 and 21.

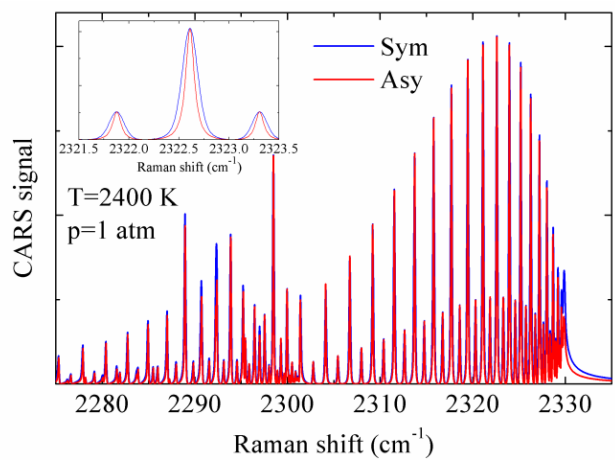

Fig. 3. Q-branch $N_2$ CARS spectrum for the symmetric Gaussian laser pulse (blue line) and the asymmetric hyperbolic secant (red line) of Fig. 1(c). The inset shows the lineshapes for $J$=19, 20 and 21.

The main differences at the bandhead and in the hot band are not removed when a large contribution from the detection equipment is present. This is shown in Fig. 4 where a broad instrument function, which makes the odd-$J$ lines disappear, is included in the calculation. Confirming the result of Fig. 3, deviations as large as 10 % are found at the bandhead and in the hot band.

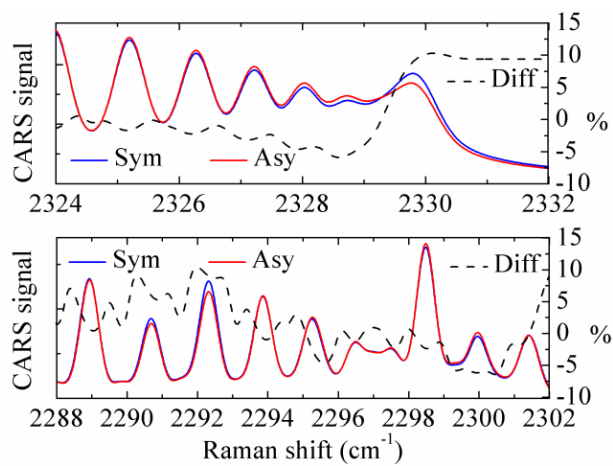

Fig. 4. Details of the Q-branch $N_2$ CARS spectrum of Fig. 3 after the incorporation of a broad instrument function of 0.6 cm$^{-1}$ against a laser bandwidth of 0.1 cm$^{-1}$. The relative difference (dashed line) is quantified on the right vertical axis.

Finally, it should be mentioned that spectral deviations between experiment and standard (Voigt) simulations of $N_2$ CARS spectra have been measured by Meyer et al. [30] who have employed ps laser pulses similar to the experimental pulse of Fig. 1 (c) [28, 29]. Thanks to the work developed in this Letter, we hypothesize a role played by the time asymmetry that was not incorporated into their calculation. Furthermore, another elaboration (not shown here) suggests an important thermometric deviation of about 100 K through ignorance of the problems posed in this work.

To conclude, we have studied an effect of line narrowing in simulated CARS lineshapes obtained by means of time asymmetric models of real ns and ps laser pulses. Apparently, the time asymmetry becomes significant in the limit of gases at low pressure when ns laser pulses are employed. Ultimately, the time asymmetry determines relevant changes in known ps-CARS spectra and its dismissal in some experimental circumstances could imply thermometric inaccuracies.

This research is promoted by the European laser research infrastructures (Laserlab Europe, proposal LLC001722, grant n. RLL3-CT-2003-506350).